\documentclass[reprint,aps,prb]{revtex4-1}
\usepackage{graphicx}
\usepackage[utf8]{inputenc}
\usepackage{amsmath}
\usepackage{bm}
\usepackage[caption=false]{subfig}
\usepackage{amsmath}
\graphicspath{{./images/}}
\usepackage{siunitx}
\usepackage{color}

\begin{document}
\title{Correlation of tunnel magnetoresistance with the magnetic properties in perpendicular CoFeB-based junctions with exchange bias}
\author{Orestis Manos$^1$, Panagiota Bougiatioti$^1$, Denis Dyck$^1$, Torsten Huebner$^1$, Karsten Rott$^1$, Jan-Michael Schmalhorst$^1$ and G\"unter~Reiss$^1$\email{Electronic mail: omanos@physik.uni-bielefeld.de}}

\affiliation{$^1$\mbox{Center for Spinelectronic Materials and Devices, Department of Physics,}\\
\mbox{Bielefeld University, Universit\"atsstra\ss e 25, 33615 Bielefeld, Germany}\\}
\date{\today}


\begin{abstract}

We investigate the dependence of magnetic properties on the post-annealing temperature/time, the thickness of soft ferromagnetic electrode and Ta dusting layer in the pinned electrode as well as their correlation with the tunnel magnetoresistance ratio, in a series of perpendicular magnetic tunnel junctions of materials sequence Ta/Pd/IrMn/CoFe/Ta$(\textit{x})$/CoFeB/MgO$(\textit{y})$/CoFeB$(\textit{z})$/Ta/Pd.  
We obtain a large perpendicular exchange bias of 79.6$\,$kA/m for $x=0.3\,$nm. For stacks with  $z=1.05\,$nm, the magnetic properties of the soft electrode resemble the characteristics of superparamagnetism.
For stacks with $x=0.4\,$nm, $y=2\,$nm, and $z=1.20\,$nm, the exchange bias presents a significant decrease at post annealing temperature $T_\textrm{ann}=330\,^{\circ}$C for 60 min, while the interlayer exchange coupling and the saturation magnetization per unit area sharply decay at $T_\textrm{ann}=340\,^{\circ}$C for 60 min. Simultaneously, the tunnel magnetoresistance ratio shows a peak of $65.5\%$ after being annealed at $T_\textrm{ann}=300\,^{\circ}$C for 60 min, with a significant reduction down to $10\%$ for higher annealing temperatures ($T_\textrm{ann}\geq330\,^{\circ}$C) and down to $14\%$ for longer annealing times ($T_\textrm{ann}=300\,^{\circ}$C for 90 min). We attribute the large decrease of tunnel magnetoresistance ratio to the loss of exchange bias in the pinned electrode. 

\end{abstract}
\maketitle

\section{Introduction}

In recent years, perpendicular magnetic tunnel junctions (p-MTJs) based on CoFeB/MgO/CoFeB have considerably advanced the development of data storage and magnetic field sensor technology. The tunnel magnetoresistance (TMR) effect in such systems\cite{Ikeda:2010} is mainly achieved by exploiting the coercivity ($H_\textrm{C}$) difference between the CoFeB electrodes, either via the growth sequence or by pinning one of them via a hard magnetic layer to a specific direction of magnetization, using the exchange bias (EB) effect \cite{Meiklejohn:1957}.\\
\mbox{ } Films exhibiting large perpendicular EB are very appealing since they provide the basis of sufficient memory retention \cite{Parkin:1999} for data storage applications i.e., the data retaining capability of the storing cell, and a wide sensing field range \cite{Nakano:2015} for sensor technology. To date, MnIr-based  \cite{ZhangIEEE:2015}, Co/Pd \cite{Lee:2016}, and Co/Pt multilayer-based synthetic antiferromagnets \cite{Kugler:2012} are commonly used as the basis for the pinned electrode of p-MTJs with EB. The TMR ratio ranging from (120-250)$\,\%$ has been reported in (Co/Pt)-based synthetic antiferromagnet pseudo-spin-valve type p-MTJs consisting of CoFeB/MgO/CoFeB \cite{Tezuka:2016,Skowronski:2017,Wang:2018}. In our previous work, we utilized MnIr and reported p-MTJs of CoFeB/MgO/CoFeB with EB around 39.8$\,$kA/m and TMR ratio of 47.2$\,\%$, at room temperature (RT) \cite{Manos}.\\ 
\mbox{ } The current work systematically investigates the correlation of magnetic properties with the measured TMR ratio in a series of p-MTJs with EB based on MnIr. Tuning the TMR ratio by altering the magnetic properties in such stacks could be desirable for the magnetic field sensor industry. Additionally, the realization of modifying the pinning properties i.e., EB, by varying the thickness of the Ta dusting layer in a MnIr-based stack, providing a useful tool for the potential fabrication of thermally-assisted MRAM in the perpendicular direction, as it is already used in the in-plane configuration \cite{Prejbeanu:2007}. In such stacks, EB-films consisted of different antiferromagnets are utilized for the pinned and soft electrode, exhibiting sufficiently different blocking temperatures and EB fields\cite{Prejbeanu:2004}.\\
\mbox{ } Huge research efforts concerning CoFeB/MgO/CoFeB magnetic tunnel junctions, unveiled the key role of some parameters such as, the MgO barrier thickness ($t_\textrm{MgO}$) \cite{Butler,Heiliger}, the thickness of the ferromagnet \cite{Yang:2015}, and the post-annealing temperature ($T_\textrm{ann}$) \cite{Hayakawa:2006,Schmalhorst:2007,Hindmarch:2010} and time \cite{Wang:2010,Wang:2011} for the optimization of TMR ratio. Nevertheless, for systems in the out-of-plane configuration the TMR ratio is inextricably linked to the degree of perpendicular magnetic anisotropy (PMA) in both electrodes, which obtains two well-defined resistance states \cite{Gan}.  
In particular, the increase of $T_\textrm{ann}$ weakens the PMA in Ta/CoFeB/MgO films \cite{Oh,Miyakawa}, resulting in a degradation of the two resistance states. The annealing effects can influence the TMR ratio in p-MTJs not only via a ``direct" channel, e.g., with the diffusion of Mn atoms to or even into the MgO barrier \cite{Hayakawa:2006,Wang:2010}, but also via an ``indirect" channel of an undesired magnetic easy axis transition. These findings indicate that a systematic investigation about the correlation of several magnetic properties with TMR ratio in such systems might offer valuable insight.\\ 
\mbox{ } We report the adjustment of the interlayer exchange coupling strength ($J$), saturation magnetization per unit area ($M_\textrm{s}t^\textrm{eff}_\textrm{FM}$), and EB from the $T_\textrm{ann}$, annealing time, $t_\textrm{MgO}$, and the thickness of the Ta interlayer $t^{\textrm{int}}_{\textrm{Ta}}$ (placed between the CoFe and CoFeB layers in the bottom electrodes). We further observe the magnetic properties of the soft electrode to resemble the characteristics of superparamagnetism for $t^{\textrm{SE}}_{\textrm{CoFeB}}=1.05$\,nm. 
The EB, $J$, and $M_\textrm{s}t^\textrm{eff}_\textrm{FM}$ show a noticeable decrease at $T_\textrm{ann}=330\,^{\circ}$C, $T_\textrm{ann}=340\,^{\circ}$C, and $T_\textrm{ann}=340\,^{\circ}$C, respectively, for annealing time equal to 60 min. In addition, we report a maximum TMR ratio of $(65.5\pm 0.5)\%$ at $T_\textrm{ann}=300\,^{\circ}$C for 60 min, realizing its significant reduction down to $(10\pm 0.3)\%$ and $(14\pm 2.3)\%$ for further increase of $T_\textrm{ann}$ or of the annealing time at $T_\textrm{ann}=300\,^{\circ}$C, respectively. 
We attribute the large reduction of TMR ratio to the EB loss which leads to the lack of antiparallel configuration between the electrodes.

\section{Preparation}

The films were deposited on thermally oxidized Si wafers at RT by DC and RF magnetron sputtering, at Ar pressure of $P=2 \cdot 10^{-3}\,$mbar. The following types of stacks were prepared and investigated
\vspace{0.5em} \\1)\mbox{ \,\,\,\,\,}Ta(4)/Pd(2)/$\textrm{Mn}_{\textrm{83}}\textrm{Ir}_{\textrm{17}}$(8)/$\textrm{Co}_{\textrm{50}}\textrm{Fe}_{\textrm{50}}$(1)/Ta$(\textit{x})$\\ \mbox{ \,\,\,\,\,\,\,\,\,\,}$\textrm{Co}_{\textrm{40}}\textrm{Fe}_{\textrm{40}}\textrm{B}_{\textrm{20}}$(0.8)/MgO(2)\\
2)\mbox{ \,\,\,\,\,}Ta(4)$\,$/$\,$Pd(2)$\,$/$\,\textrm{Mn}_{\textrm{83}}\textrm{Ir}_{\textrm{17}}$(8)$\,$/$\,\textrm{Co}_{\textrm{50}}\textrm{Fe}_{\textrm{50}} $(1)$\,$/$\,$Ta(0.3)$\,$\ 
\mbox{ \,\,\,\,\,\,\,\,\,\,}$\textrm{Co}_{\textrm{40}}\textrm{Fe}_{\textrm{40}}\textrm{B}_{\textrm{20}}$(0.8)$\,$/$\,$MgO$(\textit{y})\,$/$\,\textrm{Co}_{\textrm{40}}\textrm{Fe}_{\textrm{40}}\textrm{B}_{\textrm{20}}$(1.2)\\\mbox{ \,\,\,\,\,\,\,\,\,\,\,}Ta(3)$\,$/$\,$Pd(3)\\
3)\mbox{ \,\,\,\,\,}Ta(4)$\,$/$\,$Pd(2)$\,$/$\,\textrm{Mn}_{\textrm{83}}\textrm{Ir}_{\textrm{17}}$(8)$\,$/$\,\textrm{Co}_{\textrm{50}}\textrm{Fe}_{\textrm{50}} $(1)$\,$/$\,$Ta(0.4)$\,$\ 
\mbox{ \,\,\,\,\,\,\,\,\,\,}$\textrm{Co}_{\textrm{40}}\textrm{Fe}_{\textrm{40}}\textrm{B}_{\textrm{20}}$(0.8)$\,$/$\,$MgO$(\textit{y})\,$/$\,\textrm{Co}_{\textrm{40}}\textrm{Fe}_{\textrm{40}}\textrm{B}_{\textrm{20}}$(z)\\\mbox{ \,\,\,\,\,\,\,\,\,\,\,}Ta(3)$\,$/$\,$Pd(3),
\vspace{+0.5em}\\
\noindent where the number in parentheses is the nominal thickness of each layer in nm, $x=(0.30-0.55)\,$nm, $y=(0.6-3.0)\,$nm, and $z=(1.05-1.30)\,$nm. Ta, Pd, $\textrm{Co}_{40}\textrm{Fe}_{40}\textrm{B}_{20}$, $\textrm{Co}_{50}\textrm{Fe}_{50}$, and $\textrm{Mn}_{83}\textrm{Ir}_{17}$, MgO films were deposited from the corresponding elemental and composite targets. The purity of all targets is 99.9$\,\%$ or higher. The series of stacks were annealed at a range of $(270-400)\,^{\circ}$C for $60\,$min in vacuum ($4 \cdot 10^{-7}\,$mbar), with a magnetic field of 517.3$\,$kA/m  applied perpendicular to the film plane, in order to achieve the required coherent (001)-textured bcc crystal structure and to induce the perpendicular EB. Additionally, the third series of stacks with $x=0.4\,$nm, $y=2\,$nm, and $z=1.2\,$nm were annealed at $300\,^{\circ}$C for ($15-90\,$)min, under the same magnetic field and vacuum conditions as previously mentioned.\\
\mbox{ } Perpendicular hysteresis loops were recorded using the magnetooptical Kerr effect (MOKE) and alternating gradient magnetometer (AGM). For simplicity, in the rest of the paper the films $\textrm{Co}_{40}\textrm{Fe}_{40}\textrm{B}_{20}$, $\textrm{Co}_{50}\textrm{Fe}_{50}$, and $\textrm{Mn}_{83}\textrm{Ir}_{17}$ will be symbolized as CoFeB, CoFe, and MnIr, respectively. In addition, the layer stacks Ta(4)$\,$/$\,$Pd(2)$\,$/$\,\textrm{Mn}_{83}\textrm{Ir}_{17}$(8)$\,$/$\,\textrm{Co}_{50}\textrm{Fe}_{50} $(1)$\,$ and Ta(3)$\,$/$\,$Pd(3) will be symbolized as ``sub" and ``cap", respectively. Circular devices with $D_\textrm{device}=(0.14-1)\,\mu$m in diameter were fabricated by electron-beam lithography and Ar-ion milling \cite{Manos}. The transport properties of the p-MTJs were measured by a conventional two-probe method with a constant dc bias voltage ($V_\textrm{bias}$). 
\section{Results and discussion}

\begin{figure}[!ht]
\centering
\includegraphics[height=6.7cm, width=\linewidth]{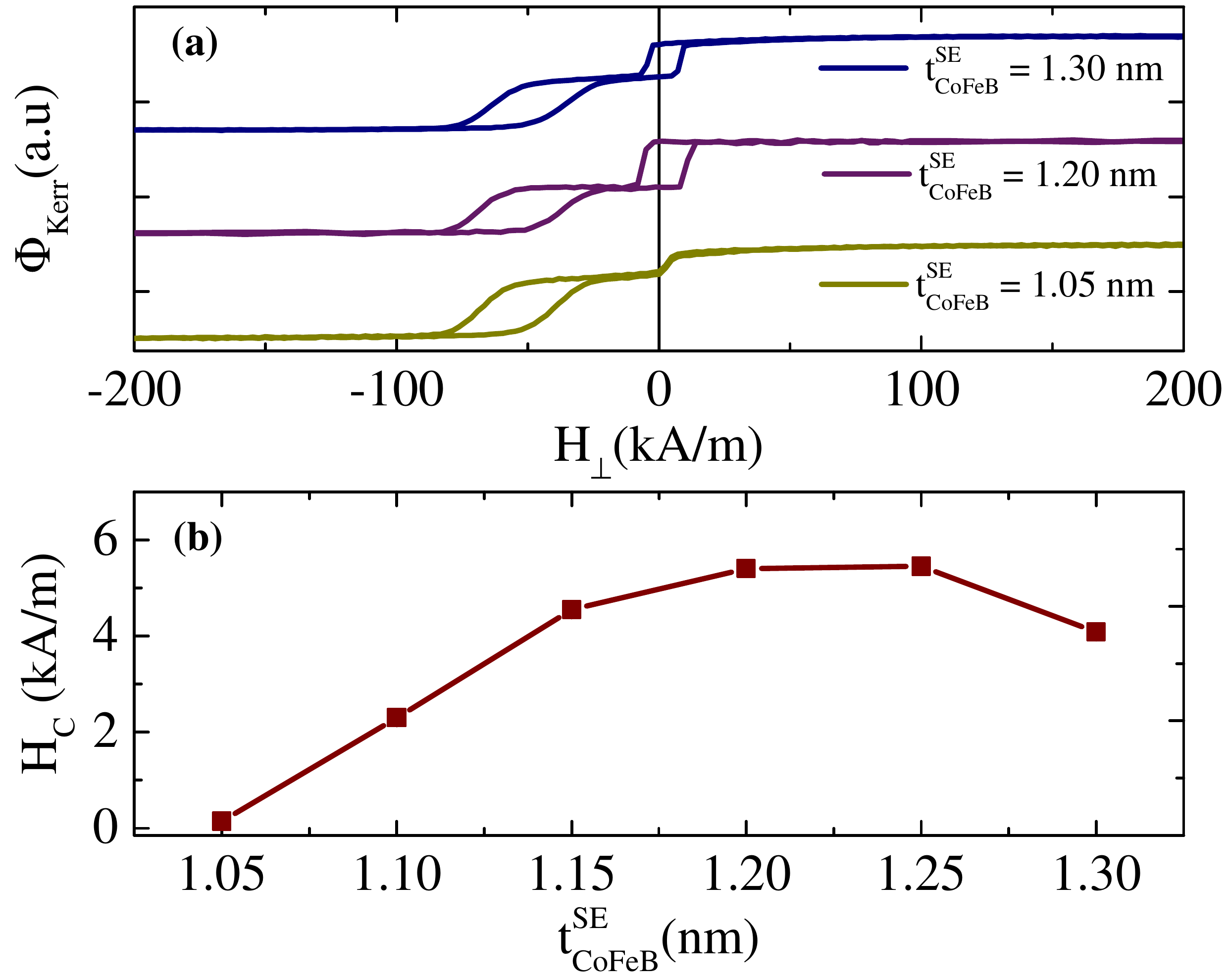}
\caption{(a) Major hysteresis loops of sub/Ta(0.4)/CoFeB(0.8)/MgO(2)/CoFeB($\textit{x}$)/cap annealed at $T_\textrm{ann}=280\,^{\circ}$C for 60 min (b) Coercivity measured in the out-of-plane direction as a function of $t^{\textrm{SE}}_{\textrm{CoFeB}}$, acquired via MOKE.}
    \label{fig:loop1}
\end{figure}

{Figure \ref{fig:loop1} (a) shows three major hysteresis loops for the stacks sub/Ta(0.4)/CoFeB(0.8)/MgO(2)/CoFeB($\textit{x}$)/cap, with $t^{\textrm{SE}}_{\textrm{CoFeB}}=$1.05 (yellow), 1.2 (purple), 1.3 (blue)$\,$nm. The two distinct magnetic steps are clearly observed arising from the corresponding soft and pinned electrodes. However, no apparent hysteresis of the soft electrode for $t^{\textrm{SE}}_{\textrm{CoFeB}}=1.05$\,nm is observed while a hysteretic behavior is evident for the rest samples with $t^{\textrm{SE}}_{\textrm{CoFeB}}=1.2$\,nm and $t^{\textrm{SE}}_{\textrm{CoFeB}}=1.3$\,nm. Figure \ref{fig:loop1}(b) presents the $H_\textrm{C}$ of the soft electrode as a function of $t^{\textrm{SE}}_{\textrm{CoFeB}}$, extracted from the minor loops (not shown). As visible, the $H_\textrm{C}$ varies in a range of $(0.1-5.4)$\,kA/m reaching its maximum for $t^{\textrm{SE}}_{\textrm{CoFeB}}=1.25$\,nm and bottoming out for $t^{\textrm{SE}}_{\textrm{CoFeB}}=1.05$\,nm. Consequently, it can be pointed out that for $t^{\textrm{SE}}_{\textrm{CoFeB}}=1.05$\,nm  the magnetic properties of the soft electrode illustrate the characteristics of superparamagnetism \cite {Bean:1959}.}

\begin{figure}[!ht]
\centering
\includegraphics[height=6.7cm, width=\linewidth]{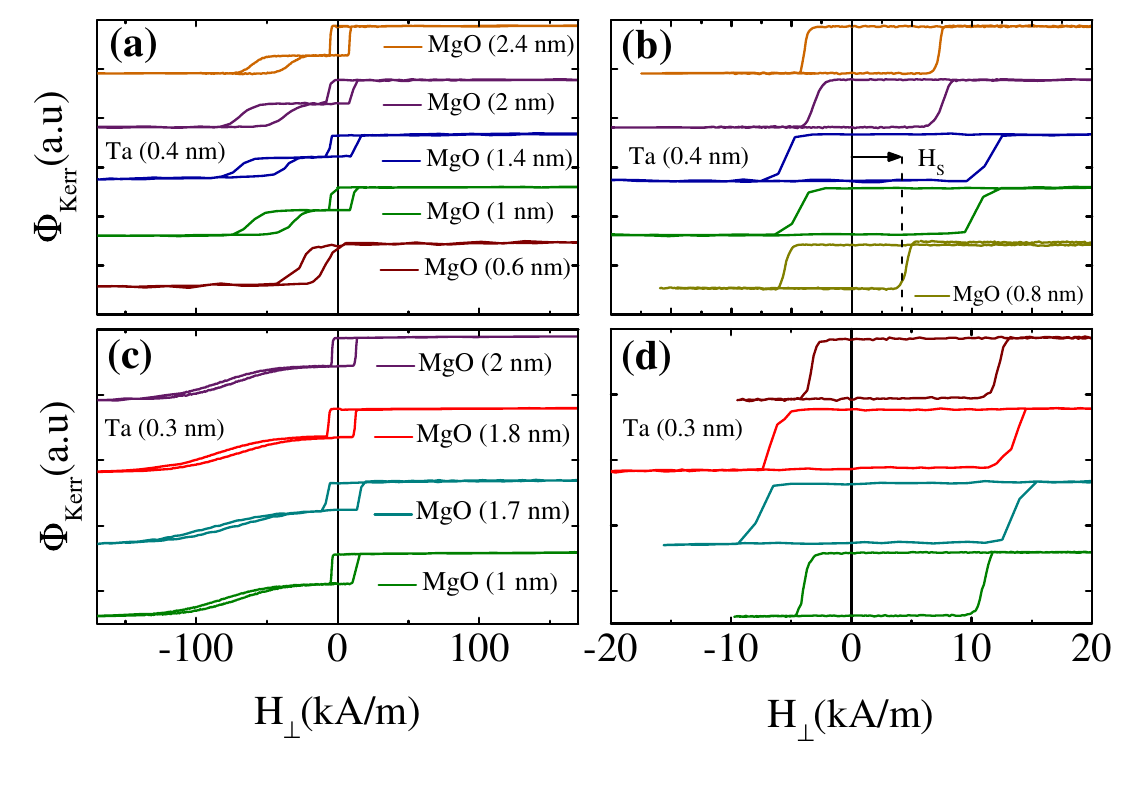}
\caption{(a), (c) Major and (b), (d) minor loops of (a), (b) sub/Ta(0.4)/CoFeB(0.8)/MgO($\textit{x}$)/CoFeB(1.2)/cap and (c), (d) sub/Ta(0.3)/CoFeB(0.8)/MgO($\textit{x}$)/CoFeB(1.2)/cap stacks, after annealing at $T_\textrm{ann}=280\,^{\circ}$C for 60 min, respectively, collected via MOKE.}
    \label{fig:loop}
\end{figure}
 
\begin{figure}[!ht]
    \centering
    \includegraphics[height=6.7cm, width=\linewidth]{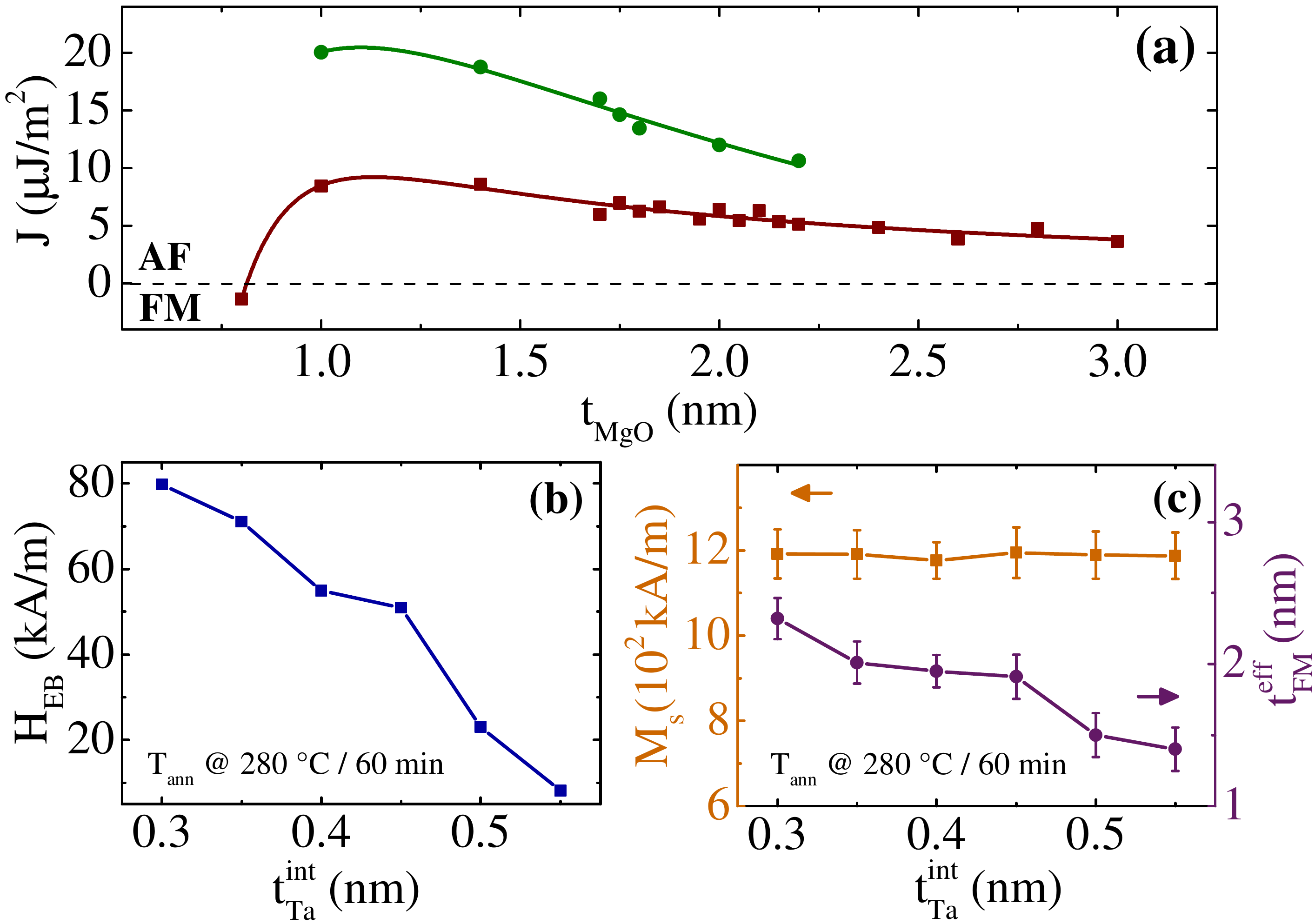}
   \caption{(a) Dependence of IEC strength ($J$) on $t_\textrm{MgO}$ with $t^{\textrm{int}}_{\textrm{Ta}}=0.4\,$nm (red squares) and $t^{\textrm{int}}_{\textrm{Ta}}=0.3\,$nm (green circles) in the EB part. The solid lines are guide to the eye. The dependence of (b) EB and (c) $M_\textrm{s}$ (left-axis), $t^\textrm{eff}_\textrm{FM}$ (right-axis) on $t^{\textrm{int}}_{\textrm{Ta}}$.}
    \label{fig:couplingvsMgo}
\end{figure}

Figure \ref{fig:loop} shows a number of representative perpendicular major and minor hysteresis loops for the stacks sub/Ta(0.4)/CoFeB(0.8)/MgO($\textit{x}$)/CoFeB(1.2)/cap (see Figs. \ref{fig:loop}(a) and (b)) and sub/Ta(0.3)/CoFeB(0.8)/ MgO($\textit{x}$)/CoFeB(1.2)/cap (see Figs. \ref{fig:loop}(c) and (d)), for $t_\textrm{MgO}=(0.6-2.4)\,$nm. A noticeable difference between both series of stacks is the enhancement of the EB field ($H_\textrm{EB}$) from 50.9$\,$kA/m to 79.6$\,$kA/m, as  $t^{\textrm{int}}_{\textrm{Ta}}$ decreases. The decrease of $t^{\textrm{int}}_{\textrm{Ta}}$ from 0.4$\,$nm to 0.3$\,$nm causes a reduction in the number of Ta interlayer atoms which are deposited on MnIr through the pinholes of CoFe sublayer, leading to the increase of $H_\textrm{EB}$ \cite{ZhangIEEE:2015}. Furthermore, as illustrated in Fig. \ref{fig:loop}(a), the two clear magnetic steps come closer to each other with decreasing $t_\textrm{MgO}$, while for $t_\textrm{MgO}=0.6\,$nm there is the formation of one magnetic step from the two initial.\\ 
\mbox{ } Moreover, the character of the interlayer exchange coupling (IEC) can be extracted by the shift ($H_\textrm{s}$) of the minor loops with respect to zero field (ferromagnetic for $H_\textrm{s}<0$ or antiferromagnetic for $H_\textrm{s}>0$) \cite{Manos}. As depicted in Figs. \ref{fig:loop}(b) and (d), the IEC character is antiferromagnetic except for the stack with $t_\textrm{MgO}=0.8$\,nm (cf. Fig. \ref{fig:loop}(b)) where a change in the character of coupling takes place, due to the ferromagnetic nature of the direct coupling between the two electrodes\cite{Vincent}. {Moritz \textrm{et al.} \cite{Moritz} suggested that in the case of films with strong PMA, the antiferromagnetic coupling can also be energetically favorable as an interplay of the magnetostatic, the exchange and anisotropy energy. The contribution of the magnetic surface charges responsible for the ferromagnetic coupling may reduce, whereas the contribution of the magnetic volume charges promoting the antiferromagnetic one is enhanced. As a result, the finally determined coupling presents an antiferromagnetic nature.}\\
\mbox{ } From the minor loops of the two series of stacks with $t^{\textrm{int}}_{\textrm{Ta}}=0.4\,$nm and $t^{\textrm{int}}_{\textrm{Ta}}=0.3\,$nm, $J$ is acquired using the formula $J=\,\mu_\textrm{0}\,H_\textrm{s}\,M_\textrm{s}\,t^\textrm{eff}_\textrm{FM}$, where $\mu_\textrm{0}$ is the permeability of free space, $M_\textrm{s}$ is the saturation magnetization, and $t^\textrm{eff}_\textrm{FM}$ is the effective ferromagnetic thickness. $t^\textrm{eff}_\textrm{FM}$ is obtained by the ferromagnetic ($t_\textrm{FM}$) and the dead layer ($t_\textrm{DL}$) thicknesses, according to the formula $t^\textrm{eff}_\textrm{FM}=t_{\textrm{FM}}-t_\textrm{DL}$. An example of the estimation of $M_\textrm{s}$ and $t_\textrm{DL}$ for the sample with $t^{\textrm{int}}_{\textrm{Ta}}=0.4\,$nm can be found in the Supplementary Material (Chap. I, including Refs. \cite{Kaidatzis,Sinha,Jang}). The $M_\textrm{s}$ and $t_\textrm{DL}$ are determined to be equal to $M_\textrm{s}=(1176\pm43)\,$kA/m ($M_\textrm{s}=(1191\pm57)\,$kA/m) and $t_\textrm{DL}=(1.05\pm0.11)\,$nm ($t_\textrm{DL}=(0.68\pm0.15)\,$nm), for the stack with $t^{\textrm{int}}_{\textrm{Ta}}=0.4\,$nm ($t^{\textrm{int}}_{\textrm{Ta}}=0.3\,$nm). The dependence of $J$ on $t_\textrm{MgO}$ for both stack series is shown in Fig. \ref{fig:couplingvsMgo}(a). As displayed in the graph there is a strong dependence of $J$ on the barrier thickness, with an antiferromagnetic (ferromagnetic) character for $t_\textrm{MgO}>0.8\,$nm ($t_\textrm{MgO}<0.8\,$nm), which is combined with an additional dependence on $t^{\textrm{int}}_{\textrm{Ta}}$. Specifically, the stacks with $t^{\textrm{int}}_{\textrm{Ta}}=0.4\,$nm show smaller values for $J$ compared to the stacks with $t^{\textrm{int}}_{\textrm{Ta}}=0.3\,$nm.\\
\mbox{ } Figures \ref{fig:couplingvsMgo}(b) and (c) depict the dependence of $H_\textrm{EB}$, $M_\textrm{s}$, and $t^\textrm{eff}_\textrm{FM}$ on $t^{\textrm{int}}_{\textrm{Ta}}$. The $M_\textrm{s}$ and $t^\textrm{eff}_\textrm{FM}$ are obtained from a series of stacks with variable CoFeB thickness (not shown). As visible from Fig. \ref{fig:couplingvsMgo}(b), there is a prominent decrease of $H_\textrm{EB}$ with increasing $t^{\textrm{int}}_{\textrm{Ta}}$. In Fig. \ref{fig:couplingvsMgo}(c) the $M_\textrm{s}$ presents a constant behaviour against $t^{\textrm{int}}_{\textrm{Ta}}$, while the $t^\textrm{eff}_\textrm{FM}$ and $M_\textrm{s}t^\textrm{eff}_\textrm{FM}$ (not shown) decrease with increasing $t^{\textrm{int}}_{\textrm{Ta}}$. Consequently, within a phenomenological approach, the higher $J$ values could be attributed to the larger magnetic torques stemming from the larger $H_\textrm{EB}$ and $t^\textrm{eff}_\textrm{FM}$ in the stacks with $t^{\textrm{int}}_{\textrm{Ta}}=0.3\,$nm.

\begin{figure}[!ht]
    \centering
    \includegraphics[height=6.7cm, width=\linewidth]{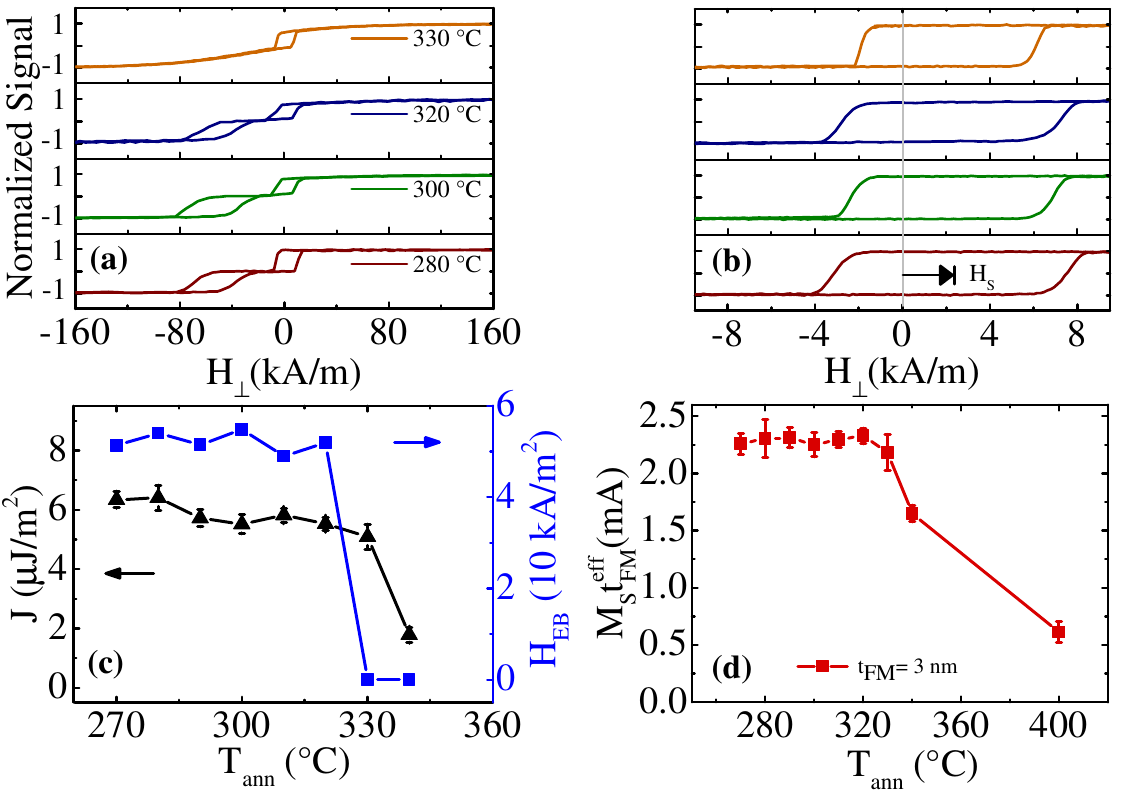}
   \caption{(a) Major and (b) minor normalized magnetic loops of stacks with $t_\textrm{MgO}=2\,$nm, $t^{\textrm{int}}_{\textrm{Ta}}=0.4\,$nm, and $t_{\textrm{CoFe+CoFeB}}=3\,$nm, for $T_\textrm{ann}$=280 (red), 300 (green), 320 (blue), 330 (orange)$\,^{\circ}$C. (c) The dependence of $J$ (left-axis) and $H_{\textrm{EB}}$ (right-axis) on $T_{\textrm{ann}}$. (d) The  $T_\textrm{ann}$ evolution of $M_\textrm{s}t^\textrm{eff}_\textrm{FM}$ for a stack with $t_\textrm{MgO}=2\,$nm, $t^{\textrm{int}}_{\textrm{Ta}}=0.4\,$nm, and $t_{\textrm{FM}}=3\,$nm.}
    \label{fig:CouplingVsannealing}
\end{figure}

\begin{figure}[!ht]
    \centering
    \includegraphics[height=6.7cm, width=\linewidth]{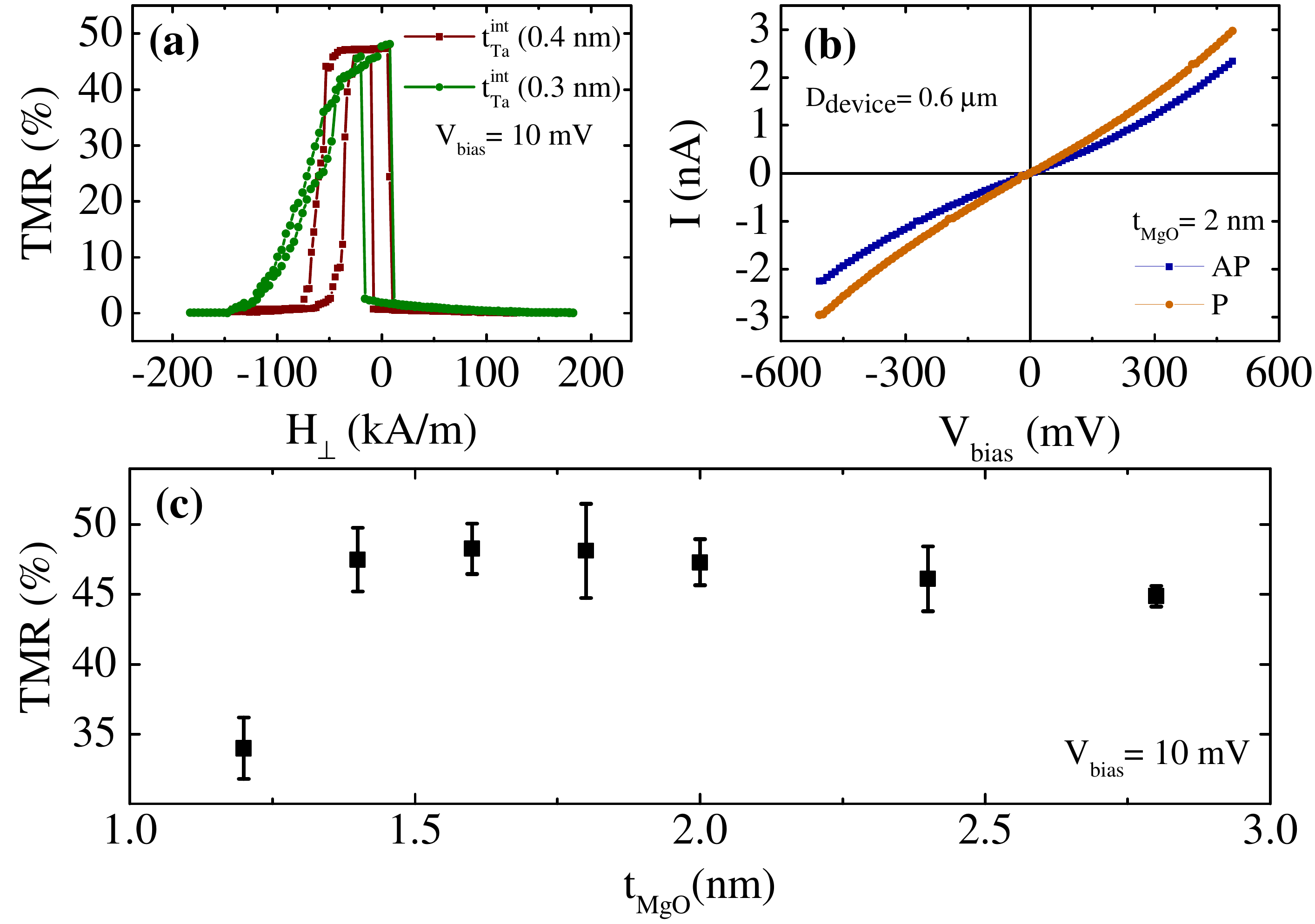}
   \caption{(a) Major TMR loops for the stacks with $t_\textrm{MgO}$=$1.8\,$nm, $t^{\textrm{int}}_{\textrm{Ta}}=0.4\,$nm (red), and $t^{\textrm{int}}_{\textrm{Ta}}=0.3\,$nm (green) in the EB part. (b) I-V characteristics for the parallel (P-orange) and antiparallel (AP-blue) states of the stack with $t_\textrm{MgO}$=$2\,$nm, $t^{\textrm{int}}_{\textrm{Ta}}=0.4\,$nm, and $D_\textrm{device}=0.6\,\mu$m. (c) TMR ratio values collected at RT with $V_\textrm{bias}=10\,$mV plotted against $t_\textrm{MgO}$.}
    \label{fig:TMRVsMgO}
\end{figure}

Figures \ref{fig:CouplingVsannealing}(a) and (b) show a number of representative normalized major and minor loops for $t_\textrm{MgO}=2\,$nm and $t^{\textrm{int}}_{\textrm{Ta}}=0.4\,$nm at several $T_\textrm{ann}$, respectively. In particular, Fig. \ref{fig:CouplingVsannealing}(a) demonstrates the loss of EB in the pinned part at $T_\textrm{ann}=330\,^{\circ}$C. Figure \ref{fig:CouplingVsannealing}(b) unveils the gradual reduction of $H_\textrm{s}$ of the free layer with increasing $T_\textrm{ann}$. In both cases, the observed behaviour can be correlated with the increased interlayer diffusion effects during post-annealing, in line with previous reports for Ta/CoFeB/MgO layer systems \cite{Miyakawa,Ikeda,Yamanouchi,Worledge} and MnIr-based CoFeB/MgO/CoFeB magnetic tunnel junctions \cite{Hayakawa:2006}.\\
\mbox{ } Figures \ref{fig:CouplingVsannealing}(c) and (d) illustrate the dependence of $J$, $H_\textrm{EB}$, and $M_\textrm{s}t^\textrm{eff}_\textrm{FM}$ on $T_\textrm{ann}$. From the minor and major loops, acquired by MOKE measurements at several $T_\textrm{ann}$, $J$ and $H_\textrm{EB}$ are calculated and presented as a function of $T_\textrm{ann}$ in Fig. \ref{fig:CouplingVsannealing}(c). As visible from the graph, the high $T_\textrm{ann}$ causes a significant degradation of $J$ (left-axis) and $H_\textrm{EB}$ (right-axis) bottoming out at $T_\textrm{ann}=340\,^{\circ}$C and $T_\textrm{ann}=330\,^{\circ}$C, respectively. Similar temperature dependent behaviour of the coupling energy density was reported by Yakushiji \textit{et al.} \cite{Yakushiji:2015}, in perpendicularly magnetized synthetic antiferromagnetically coupled reference structures. Furthermore, the evolution of $J$ and $H_\textrm{EB}$ with the annealing time can be found in the Supplementary Material \cite{Manos:Supp} (Chap. II), where the loss of EB for 90 min annealing time and a slight decrease of $J$ for annealing time$\geq$60 min can be extracted.\\
\mbox{ } Additionally, in Fig. \ref{fig:CouplingVsannealing}(d) the $M_\textrm{s}t^\textrm{eff}_\textrm{FM}$ presents stable values for $270\,^{\circ}$C$\,\leq T_{\textrm{ann}}\leq330\,^{\circ}$C noting a strong decrease for $T_{\textrm{ann}}\geq340\,^{\circ}$C, reaching a low at $T_{\textrm{ann}}=400\,^{\circ}$C. In order to further investigate the influence of diffusion effects on the magnetic properties of the stack, magnetic measurements are performed in a series of films with variable thickness of the top CoFeB, at various $T_{\textrm{ann}}$. Further analysis and discussion can be found in the Supplementary Material \cite{Manos:Supp} (Chap. I, including Refs. \cite{Kaidatzis,Sinha,Jang}).\\
\mbox{ } Figure \ref{fig:TMRVsMgO}(a) depicts two representative major TMR loops for a series of stacks with  $t^{\textrm{int}}_{\textrm{Ta}}=0.3\,$nm (green) and $t^{\textrm{int}}_{\textrm{Ta}}=0.4\,$nm (red), for $t_\textrm{MgO}=2\,$nm at $V_{\textrm{bias}}= 10\,$mV with $D_\textrm{device}=0.6\,\mu$m. In Fig. \ref{fig:TMRVsMgO}(b), the I-V curves of the stack with $t_\textrm{MgO}=2\,$nm and $t^{\textrm{int}}_{\textrm{Ta}}$=$0.4\,$nm are illustrated for the parallel (orange) and the antiparallel (blue) magnetic alignment of both electrodes. From a series of I-V curves for stacks with $t_\textrm{MgO}=(1.2-2.8)\,$nm, the $\textrm{TMR}$ ratio equal to  $\textrm{TMR}=\frac{I_\textrm{P}-I_\textrm{AP}}{I_\textrm{AP}}$ can be extracted, where $I_\textrm{P}$ ($I_\textrm{AP}$) is the current in the parallel (antiparallel) state. Figure \ref{fig:TMRVsMgO}(c) displays the averaged TMR ratio extracted from 8 devices at $V_\textrm{bias}=10\,$mV, acquired from the I-V curves, plotted against $t_\textrm{MgO}$ with $D_\textrm{device}=0.6\,\mu$m. The TMR ratio increases with the $t_\textrm{MgO}$, reaching a saturation for $t_\textrm{MgO}\geq1.4\,$nm with a slight decrease for large $t_\textrm{MgO}$. In a phenomenological approach, the TMR ratio increase with $t_\textrm{MgO}$ can be attributed to the increase of the tunnel probability for electrons with an off-normal incidence, which results in an increase of the effective polarization of the tunnel current and, therefore, the measured TMR ratio \cite{Yuasa:2004}. 

\begin{figure}[!ht]
    \centering
    \includegraphics[height=7cm, width=\linewidth]{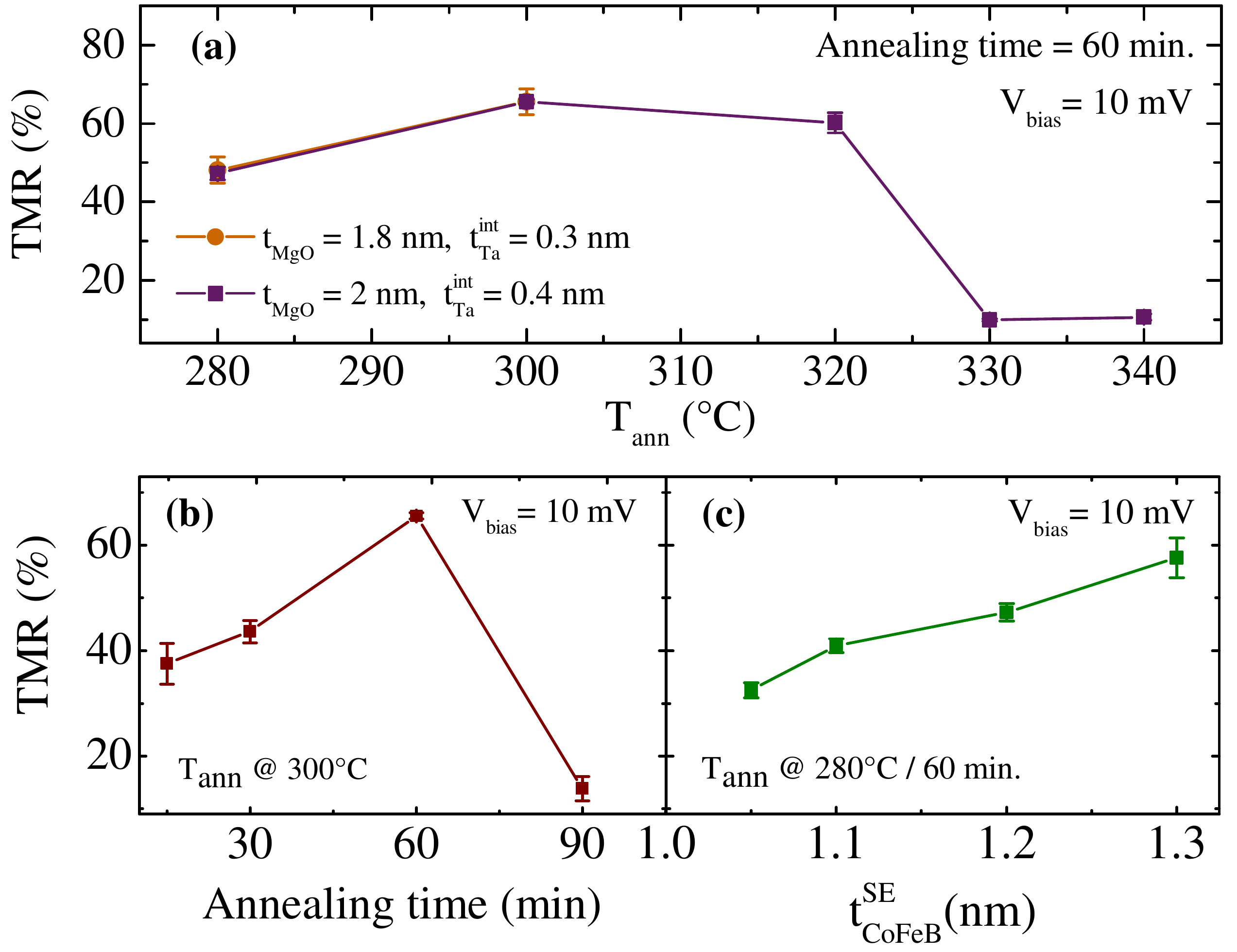}
   \caption{TMR ratio dependence on (a) $T_\textrm{ann}$, (b) annealing time, (c) $t^{\textrm{SE}}_{\textrm{CoFeB}}$ for the stacks having $D_\textrm{device}=0.6\,\mu$m with $t_\textrm{MgO}=1.8\,$nm, $t^{\textrm{int}}_{\textrm{Ta}}=0.3\,$nm (circles) and $t_\textrm{MgO}=2\,$nm, $t^{\textrm{int}}_{\textrm{Ta}}=0.4\,$nm (squares).}
    \label{fig:TMRvsAnnealing}
\end{figure}

Figure \ref{fig:TMRvsAnnealing} illustrates the dependence of TMR ratio on the $T_\textrm{ann}$ for annealing time equal to 60 min, annealing time for $T_\textrm{ann}=300\,^{\circ}$C, and $t^{\textrm{SE}}_{\textrm{CoFeB}}$ extracted from the I-V characteristics, for the stacks with $t_\textrm{MgO}=1.8\,$nm, $t^{\textrm{int}}_{\textrm{Ta}}=0.3\,$nm (circles) and $t_\textrm{MgO}=2\,$nm, $t^{\textrm{int}}_{\textrm{Ta}}=0.4\,$nm (squares). Specifically, in Fig. \ref{fig:TMRvsAnnealing}(a) the TMR ratio initially increases with increasing the $T_\textrm{ann}$ reaching a maximum of $(65.5\pm 3.2)\%$ ($(65.5\pm 0.5)\%$) at $T_\textrm{ann}=300\,^{\circ}$C for the sample series with $t^{\textrm{int}}_{\textrm{Ta}}=0.3\,$nm ($t^{\textrm{int}}_{\textrm{Ta}}=0.4\,$nm), presenting $H_\textrm{EB}=79.6\,$kA/m  ($H_\textrm{EB}=50.9\,$kA/m). At $T_\textrm{ann}=320\,^{\circ}$C there is a gradual decrease of TMR ratio of $60\,\%$ possibly attributed to the existence of Mn atoms in the MgO barrier \cite{Hayakawa:2006}. For $T_\textrm{ann}\geq330\,^{\circ}$ a steep reduction of TMR ratio can be observed. Moreover, a similar trend is presented in Fig. \ref{fig:TMRvsAnnealing}(b) with increasing TMR ratio for increasing the annealing time, peaking at 60 min and bottoming out at 90 min. The observed behaviours of increasing TMR ratio with increasing the $T_\textrm{ann}$ and annealing time result from the crystallization of the amorphous CoFeB electrodes and the improvement of crystalline structure of MgO (001) barrier \cite{Almasi:2017}.\\
\mbox{ } Taking into account the $T_\textrm{ann}$ and annealing time dependence of $J$, EB, TMR, and $M_\textrm{s}t^\textrm{eff}_\textrm{FM}$ (only $T_\textrm{ann}$ dependence), the steep reduction of TMR ratio at $T_\textrm{ann}=330\,^{\circ}$C for 60 min and $T_\textrm{ann}=300\,^{\circ}$C for 90 min annealing time, coincides with the EB loss at these specific conditions. Therefore, among these three magnetic parameters the EB appears to have the most important influence on TMR. The EB loss leads to the lack of antiparallel configuration between the electrodes, which is necessary for the establishment of two well-defined resistance states. The aforementioned behaviour is also reported by Gan \textit{et al.} \cite{Gan} in CoFeB-based p-MTJs, where the lack of antiparallel configuration originates from the different temperature dependence of the $H_\textrm{C}$ of the individual electrodes.\\
\mbox{ } In Fig. \ref{fig:TMRvsAnnealing} (c) the monotonic increase of TMR ratio with increasing the $t^{\textrm{SE}}_{\textrm{CoFeB}}$ can be extracted. Specifically, the TMR ratio is equal to $(32.5\pm 1.4)\%$ for $t^{\textrm{SE}}_{\textrm{CoFeB}}=1.05$\,nm taking its highest value of $(57.6\pm 3.8)\%$ for $t^{\textrm{SE}}_{\textrm{CoFeB}}=1.30$\,nm. The enhancement of the TMR ratio with increasing the $t^{\textrm{SE}}_{\textrm{CoFeB}}$ could be interpreted as an outcome of the enhanced spin polarization of the d[001] states as Yang \textit{et al.} reported in their work \cite{Yang:2015}. As an example in Fig. \ref{fig:loop1}(a), for $t^{\textrm{SE}}_{\textrm{CoFeB}}=1.05$\,nm the soft electrode follows a superparamagnetic behaviour which results in a significant weaking of spin polarization leading to small TMR ratio values.

\vspace{0.5em}
\section{Conclusion}
In conclusion, we have systematically studied the correlation of magnetic properties with the measured TMR ratio on a series of p-MTJs with EB of materials sequence Ta/Pd/IrMn/CoFe/Ta/CoFeB/MgO/CoFeB/Ta/Pd. We reported the modulation of $J$, $M_\textrm{s}t^\textrm{eff}_\textrm{FM}$, and EB from the parameters $T_\textrm{ann}$, annealing time, $t^\textrm{int}_\textrm{Ta}$ in the pinned electrode. For $t^\textrm{SE}_\textrm{CoFeB}=1.05$\,nm the magnetic properties of the soft electrode showed the characteristics of superparamagnetism. A strong dependence of $J$, $M_\textrm{s}t^\textrm{eff}_\textrm{FM}$, and EB on $t^{\textrm{int}}_{\textrm{Ta}}$ and $T_\textrm{ann}$ was pointed out, reporting $H_\textrm{EB}=79.6\,$kA/m for $t^{\textrm{int}}_{\textrm{Ta}}=0.3\,$nm. 
In addition, 
after the variation of $T_\textrm{ann}$ and annealing time we realized a TMR ratio in the range of $(10-65.5)\%$ noting a steep reduction of TMR ratio at $T_\textrm{ann}=330\,^{\circ}$C for 60 min and $T_\textrm{ann}=300\,^{\circ}$C for 90 min. Comparing the $T_\textrm{ann}$ and annealing time dependence of $J$, EB, TMR, and $M_\textrm{s}t^\textrm{eff}_\textrm{FM}$ (only $T_\textrm{ann}$ dependence), we conclude that the loss of EB is the major factor for the large decrease of TMR ratio in the examined stacks. Controlling the TMR ratio by changing the magnetic properties in such stacks is of great interest for the magnetic field sensor industry.  

 \section{Supplementary Material}
See supplementary material for further information about the magnetic measurements at several post annealing temperatures and the evolution of interlayer exchange coupling and EB with the annealing time.

\section{Acknowledgment}
The authors gratefully acknowledge the financial support by the Deutsche Forschungsgemeinschaft (DFG, contract RE1052/22-2).

\bibliographystyle{apsrev4-1}
\bibliography{main}

\end{document}